\documentclass[a4paper]{VisionStyle}
\usepackage{epsfig}

\renewcommand{\tabcolsep}{2pt}

\begin{document}

\title{Ultraluminous X-ray Sources in M51}

\author{
Y.\,Terashima\inst{1,2}
\and
A.S.\,Wilson\inst{1,3}
} 

\institute{
Astronomy Department, University of Maryland, College Park, MD 20742, USA
\and 
Institute of Space and Astronautical Science, 3-1-1 Yoshinodai,
 Sagamihara, Kanagawa 229-8510, Japan
\and 
Space Telescope Science Institute, 3700 San Martin Drive, Baltimore, MD 21218, 
USA
}

\maketitle 

\begin{abstract}

We present the results of {\it Chandra} observations of off-nuclear
X-ray sources in the spiral galaxy M51 (NGC 5194). 113 X-ray sources
have been detected in the field of view and 84 among them project
within the disk of NGC 5194. Six and twenty eight sources have
luminosities exceeding $1\times10^{39}$ ergs s$^{-1}$ and
$1\times10^{38}$ ergs s$^{-1}$ in the 0.5--8 keV band, respectively.
The number of luminous sources is much higher than normal spiral and
elliptical galaxies and similar to galaxies experiencing starburst
activity.  X-ray spectra of most of the detected sources are
consistent with a power law with a photon index between 1 and 2, while
one source has an extremely hard spectrum and two sources have soft
spectra.  The spectra of three ultraluminous sources are consistent
with both a power law and a multicolor disk blackbody (MCD) model,
while a power law model is preferable to a MCD model in two
objects. One luminous object which shows a remarkable spectrum,
including emission lines, is also found and discussed.

\keywords{galaxies: active 
--- X-rays: galaxies --- galaxies: Individual (M51)}
\end{abstract}

\section{Introduction}

X-ray emission from spiral galaxies consists of various components,
namely discrete sources such as X-ray binaries, hot gas, an active
galactic nucleus, if present, and so on. One of the most intriguing
and puzzling classes of X-ray sources in spiral galaxies is
ultraluminous compact X-ray sources (ULXs) whose luminosities
($>10^{39}$ ergs s$^{-1}$) well exceed the Eddington luminosity of
neutron stars.

  The nearby (8.4 Mpc; Feldmeier et al. 1997) spiral galaxy NGC 5194
(M51) is one of the best targets to study an X-ray source population
including ULXs since previous observations with the {\it ROSAT} PSPC and
HRI have shown the presence of several luminous X-ray sources whose
luminosities exceed $10^{39}$ ergs s$^{-1}$ (Marston et al. 1995;
Ehle, Pietsch, \& Beck 1995; Roberts \& Warwick 2000). In this paper,
we present results of two {\it Chandra} observations and discuss the
source population, time variability, and spectra of luminous X-ray
sources. The detailed results can be found in Terashima \& Wilson
(2002).

\section{Observations and Source Detection}

{\it Chandra} observations of M51 were performed on 2000 June 20 and
2001 June 23 with the ACIS-S3 chip, with effective exposure times
of 14.9 ksec and 26.8 ksec, respectively. 

A true color image is shown in Fig. 1. In this image, we see extended
emission distributed along the spiral arms of the host galaxy of NGC
5194, a bright nuclear region, and many discrete sources.
The companion galaxy NGC 5195 also show discrete sources and
diffuse emission. The color indicates that the diffuse emission has
a very soft spectrum, while the discrete sources show a variety of spectral
hardness.  We performed X-ray source detection using wavdetect in the
CIAO software package for the first, second, and combined data sets,
and detected 113 source in the field of view.  84 and 12 objects are
spatially coincident with NGC 5194 and NGC 5195,
respectively. According to the $\log N - \log S$ derived from {\it
Chandra} deep surveys, 8--12 background sources are expected in the
field of view of the S3 chip. Most of the discrete sources in NGC 5194
are located on or near the spiral arms and few sources are seen
between the arms.

We used the combined data to measure the luminosity function of the
X-ray sources in NGC 5194. Band ratios (2$-$8 keV / 0.5$-$2 keV) were
used to estimate the spectral shapes and to calculate source
luminosities (Fig. 2). X-ray spectra of most sources are consistent
with a phton index between 1 and 2 if a power law spectrum modified by
the Galactic absorption is assumed. The luminosity function is shown
in Fig. 3. The slope in the luminosity range above $10^{38}$ ergs
s$^{-1}$, which is virtually free from incompleteness, is --0.89. This
slope is in the range often observed in disk galaxies and starburst
galaxies, and significantly flatter than those observed in ellipticals
and lenticulars.

Next we searched for short-term and long-term variability.
Within each observation, three sources
were found to be time variable on a time scale of several 1000 s.
About one-third (26) of discrete sources in NGC 5194 indicate time
variability at greater than the 3$\sigma$ level in the 0.5$-$8 keV band
between the two observations separated a year. Four objects among them
show a large amplitude (a factor of ten or more) variability suggesting their
transient nature.

\label{fauthor-E1_sec:fig}

\begin{figure}[ht]
  \begin{center}
    \epsfig{file=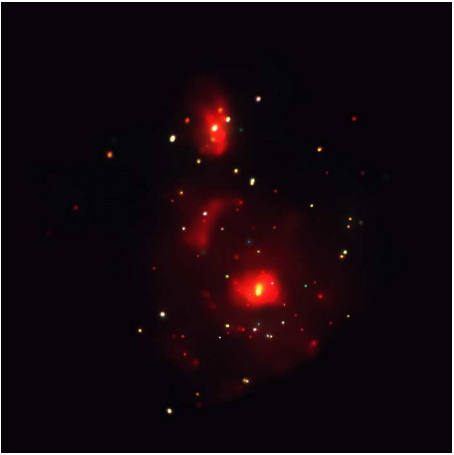, width=6cm}
  \end{center}
\caption{Adaptively smoothed true color image. 
Red: 0.3--1.5 keV, Green: 1.5--3 keV, and Blue: 3-8 keV.}  
\label{yterashima-E3_fig:fig1}
\end{figure}

\begin{figure}[ht]
  \begin{center}
    \epsfig{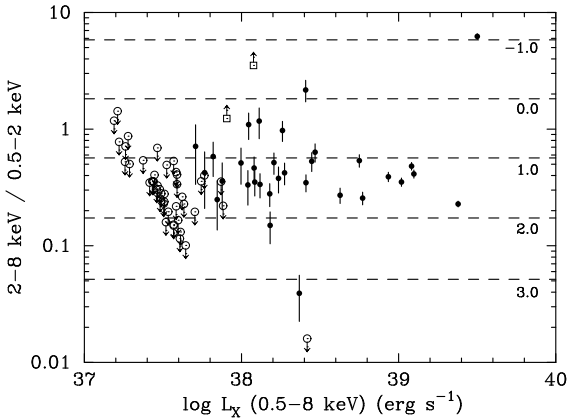}
  \end{center}
\caption{Luminosity dependence of the band ratio (2--8 keV)/(0.5--2 keV) for 
the X-ray sources in NGC 5194. Horizontal dashed lines corresponds to photon
indices of -1, 0, 1, 2, and 3 when a power law model absorbed by the Galactic
column is assumed. Open squares and open circles represent the objects 
detected only in the hard band or soft band, respectively.}
\label{yterashima-E3_fig:fig2}
\end{figure}

\begin{figure}[ht]
  \begin{center}
    \epsfig{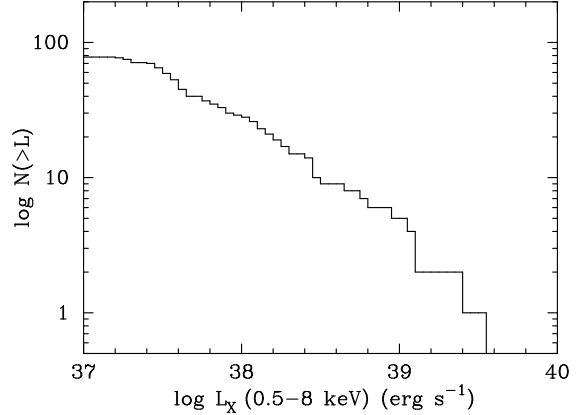}
  \end{center}
\caption{Luminosity function of the X-ray source in NGC 5194.}  
\label{yterashima-E3_fig:fig3}
\end{figure}

\begin{table}[thb]
	\caption{Spectral Fit to Ultra Luminous X-ray Sources}
  \label{fauthor-E1_tab:tab1}
  \begin{center}
    \leavevmode
    \footnotesize
    \begin{tabular}[h]{lcccccl}
      \hline \\[-5pt]
Source & $N_{\rm H}^b$	& $\Gamma$	& $kT_{\rm in}^c$	& $\chi^2$/dof	& $L_{\rm X}^d$\\
Name$^a$& 			&	& 	&		& \\
	\hline \\[-5pt]
5 (1)	& $0.18\pm0.11$ 	& $1.37^{+0.38}_{-0.28}$& ...	& 11.8/12	& 1.2\\
	& $0.099^{+0.10}_{-0.069}$& ...& $1.85^{+0.85}_{-0.74}$ & 10.6/12	& 1.2\\
5 (2)	& 0 ($<$0.073)		& $1.17^{+0.40}_{-0.26}$& ... 	& 12.3/8	& 0.55\\
	& 0 ($<0.038$)		& ...& $1.58^{+1.13}_{-0.55}$	& 13.6/8	& 0.41\\
26 (1)	& $5.8^{+6.9}_{-4.8}$	& $1.8^{+2.1}_{-1.1}$	& ...	& 3.2/3		& 2.9\\
26 (2)	& $3.6^{+2.0}_{-0.7}$	& $1.55^{+0.86}_{-0.60}$& ...	& 12.2/14	& 4.0\\
37 (2)	& $0.11^{+0.04}_{-0.05}$& $1.55^{+0.19}_{-0.15}$& ...	& 25.0/33	& 1.5\\
	& 0 ($<$0.023)		& ...& $1.76^{+0.29}_{-0.37}$	& 35.9/33	& 1.3\\
41 (1)	& 0.11 ($<$0.18)	& $1.50^{+0.38}_{-0.22}$& ...	& 6.8/11 	& 1.1\\
	& 0.018 ($<$0.11)	& ...	& $1.43^{+0.67}_{-0.40}$& 5.5/11 	& 0.84\\
41 (2)	& $0.090^{+0.050}_{-0.057}$& $1.32\pm0.17$	& ...	& 27.1/23 	& 1.1\\
	& 0.0012 ($<$0.042)	& ...	& $2.20^{+0.68}_{-0.49}$& 26.6/23	& 1.0\\
69 (1)	& $0.12^{+0.03}_{-0.06}$& $1.24^{+0.12}_{-0.17}$& ...	& 34.9/28	& 2.7\\
	& 0.034 ($<$0.073)	& ...	& $2.34^{+0.97}_{-0.53}$& 37.0/28	& 2.4\\
69 (2)	& $0.83^{+0.81}_{-0.51}$& ...	& $0.17^{+0.13}_{-0.06}$& 3.3/6$^e$	& 0.56\\
82 (1)& $0.16\pm0.05$	& $2.26^{+0.24}_{-0.21}$& ...		& 52.7/47 	& 3.1\\
	& 0 ($<$0.021)	& ...	& $0.87\pm0.09$			& 60.5/47 	& 2.1\\
82 (2)	& $0.097^{+1.9}_{-0.041}$ & $1.86^{+0.19}_{-0.17}$ & ...& 66.0/47	& 1.8\\
	& 0			& ...		& 1.11		& 91.9/47	& ...\\
      \hline \\
      \end{tabular}
  \end{center}
({\it a}): (1) and (2) denote the first and second observations, respectively.
({\it b}): Absorption column density in unit of $10^{22}$ cm$^{-2}$.
({\it c}): Inner temperature for multicolor disk blackbody model in unit of keV.
({\it d}): X-ray luminosity in the 0.5--8 keV band in unit of $10^{39}$ ergs s$^{-1}$ corrected for absorption.
({\it e}): Maximum-likelihood fit using C-statitic.
\end{table}

\section{Ultra Luminous X-ray Sources}

Spectral fits were performed for relatively bright objects. Here we
present the results for six ULXs with 0.5--8 keV luminosities exceeding
$10^{39}$ ergs s$^{-1}$ in at least one observation.  Two models - a
power law and a multicolor disk blackbody (MCD) - were examined. The
results are summarized in Table 1. The errors are 90\% confidence for
one parameter of interest ($\Delta \chi^2=2.7$).

\subsection{ULXs with a Power law Spectrum}

The spectra of NGC 5194 \#82 = CXOM51 J133007.6 +471106 and NGC 5194
\#37 = CXOM51~J132953.3 \\+471042 in the second observation are better
fitted with a power law than a MCD (the source names are taken from
Terashima \& Wilson 2002). The {\it ASCA} spectra of ULXs studied by
Makishima et al. (2000) are successfully explained by a MCD rather
than a power law, while ULXs whose spectra are better fitted by a
power law have been found in recent observations (sources in IC 342,
Kubota et al. 2001; CXOU J112015.8+133514 in NGC 3628, Strickland et
al. 2001; in NGC 4565 (RXJ 1236.2+2558 = XMM J123617.5+285855,
Foschini et al. 2002; NGC 5204 X-1, Roberts et al. 2001). Thus it
appears that these objects constitute a class of ULXs whose spectra
are a power law.

The power law spectrum may be accounted for in several models: (1)the
``hard'' state of a black hole, in which a power law spectrum with a
photon index of 1.4--2 (e.g., Tanaka 1996, Done 2001) is found, (2)
beamed X-ray emission in ``micro quasars'' (K\"ording, Falcke, \&
Markoff 2002), and so on.

NGC 5194 \#37 is detected only in the second observations which
indicates the photon flux increased by a factor of more than 100 in
the 0.5$-$8 keV band.

On the other hand, NGC 5194 \#82 is detected and spectral information
is available in both observations. The spectrum flattened ($\Gamma =
2.26$ to 1.86) accompanied by a flux decrease by 40\%. This behavior
is similar to that of Galactic BHCs in the hard state which have
steeper spectra in brighter-flux states.

\subsection{Spectral Steepening in NGC 5194 \#69}

NGC 5194 \#69 = CXOM51~J133001.0+471344 shows a remarkable spectral
variability (Fig. 4). The photon flux in the 2--8 keV band decreased
by a factor of more than 13.  The first spectrum is relatively hard, with a
photon index $\Gamma = 1.24^{+0.12}_{-0.17}$ or $kT_{\rm
in}=2.3^{+1.0}_{-0.5}$ keV, while the second spectrum is extremely
soft, with $\Gamma>5.1$, $kT_{\rm in}=0.17^{+0.13}_{-0.06}$ keV
(MCD), or $kT = 0.16^{+0.08}_{-0.06}$ keV (black body).  If we adopt a
power law model and a MCD model for the first and second spectra,
respectively, the intrinsic luminosity of this source declined by a
factor of 3500 in the 2--10 keV bands.

Such a drastic steepening of spectra accompanied by a large flux
decline is seen in soft X-ray transients which show very soft spectra
($kT=0.2-0.3$ keV for a black body model) when the luminosity goes
down below $10^{34}$ ergs s$^{-1}$ (Tanaka 1996, Tanaka \& Shibazaki
1996).  Such luminosities, however, are much lower than we observed in
the source NGC 5194 \#69 in the second observation ($5.6\times10^{38}$
ergs s$^{-1}$ in 0.5--8 keV and $7.4\times10^{35}$ ergs s$^{-1}$ in
2--10 keV). Galactic black hole candidates are in a hard state, with
spectral described by a power law with a photon index of 1.5--2.0, for
such a luminosity.  Thus, it is not clear whether the similarity of
the spectral shape between the quiescent state of soft X-ray
transients and the NGC 5194 source in the second observation has any
physical significance.

\begin{figure}[ht]
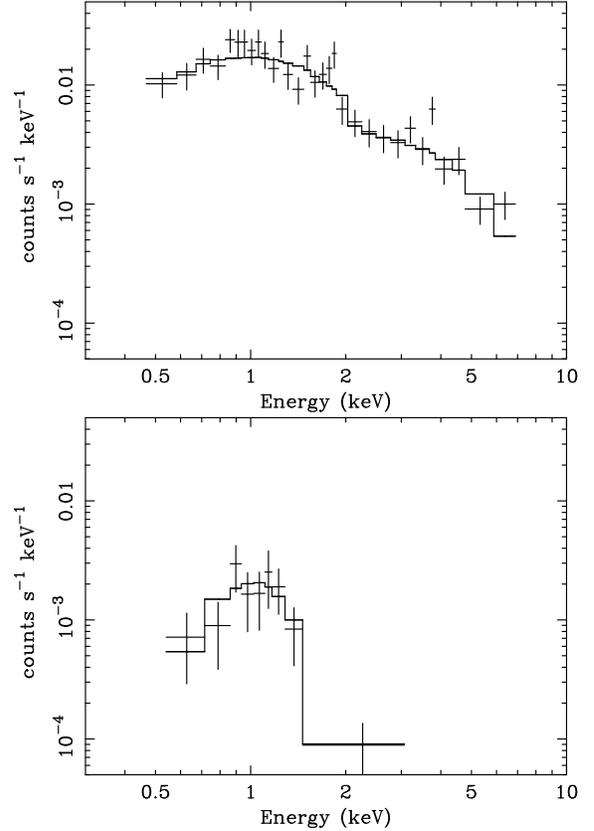

  \begin{center}
    \epsfig{file=yterashima-E3_fig4a.ps, width=5.5cm, angle=-90}
    \epsfig{file=yterashima-E3_fig4b.ps, width=5.5cm, angle=-90}
  \end{center}
\caption{Chandra spectra of NGC 5194 \# 69 in the first (top) and second 
(bottom) observation.}  
\label{yterashima-E3_fig:fig4}
\end{figure}

\subsection{The Hard Source NGC 5194 \#26}

  One source (CXOM51~J 132950.7+471155 = NGC 5194 \#26) has 
the hardest spectrum among the detected sources. This source can
been seen as a very blue (= very hard) source to the NW of the nucleus
in Fig. 1 of Terashima \& Wilson
(2001).

  This source shows a remarkable spectrum and variability. In the
first observation, the spectrum is apparently flat and virtually no
photons are detected in the low energy band below 2 keV. Since there
are indications of emission lines, we added four Gaussians to
represent these features to an absorbed power law continuum
($\Gamma=1.6$, $N_{\rm H}=4.6\times10^{22}$ cm$^{-2}$) (Fig. 5
top). The intrinsic luminosity is $3.4\times10^{39}$ ergs s$^{-1}$ in
the 0.5$-$8 keV band.  On the other hand, emission line features are
not clear in the spectrum of the second observation (Fig. 5
bottom). The continuum is fitted with a partially covered power law
model ($\Gamma=1.8$, $N_{\rm H}=4.5\times10^{22}$ cm$^{-2}$, covering
fraction = 0.988). The intrinsic luminosity ($5.0\times10^{39}$ ergs
s$^{-1}$ in 0.5--8 keV) is 50\% higher than that in the first observation.

Emission lines with large equivalent widths from highly ionized atoms
are observed in eclipses of high-mass X-ray binaries, when the direct
emission from the compact object is blocked by the companion star, and
only emission from and scattered by the photoionized medium
surrounding the system (such as a stellar wind from the companion) is
observed.  However, the observed luminosity is too large for such
scattered emission from ordinary binaries. This dichotomy could be
reconciled if the compact object is an ULX or the irradiating emission
is anisotropic. The spectrum without strong emission lines and with a larger
luminosity in the second observation appears to be dominated not by
scattered emission but by direct emission from the compact object and
suggests that the possibility of beaming to explain the emission lines
in the first observation.

\begin{figure}[ht]
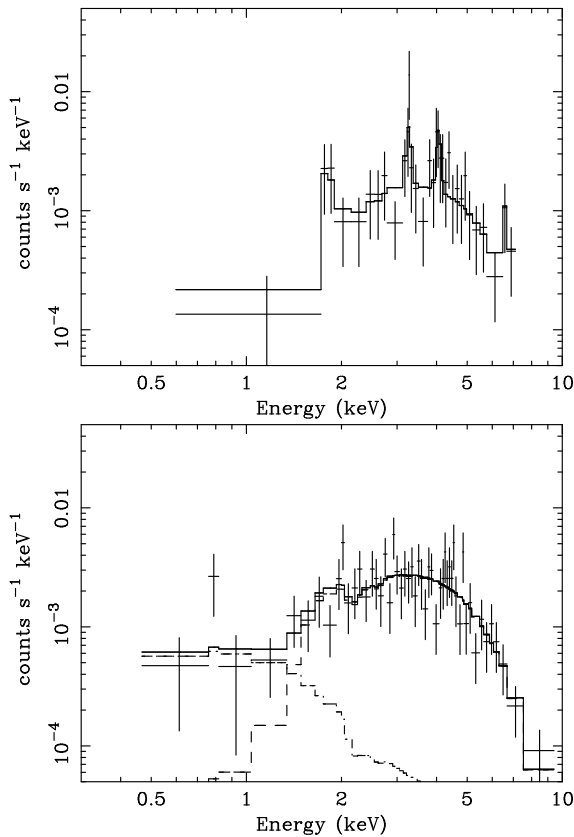

  \begin{center}
    \epsfig{file=yterashima-E3_fig5a.ps, width=5.5cm, angle=-90}
    \epsfig{file=yterashima-E3_fig5b.ps, width=5.5cm, angle=-90}
  \end{center}
\caption{Chandra spectra of NGC 5194 \# 26 in the first (top) and second 
(bottom) observation.}  
\label{yterashima-E3_fig:fig5}
\end{figure}

\subsection{Two Other ULXs}

The spectra of the two ULXs NGC 5194 \#5 = CXOM51 J132939.5+471244 and
NGC 5194 \#41 = CXOM51\\J132953.7 +471436 can be represented by either
a power law or a MCD model. The luminosity of the former source
decreased by a factor of 2.5, while that of the latter remained the same
between the two observations. Their spectra flattened.  The trend of
the spectral variability of NGC 5194 \#5 (flatter in lower flux state)
is similar to both the hard state of Galactic black hole candidate,
and ULXs whose spectra are described by a MCD model (Mizuno, Kubota,
\& Makishima 2001).  The spectral slope for the power law fit and
$kT_{\rm in}$ for the MCD fit are in the range for previously known
ULXs.\\

\begin{acknowledgements}


Y.~T. is supported by the Japan Society for the Promotion of Science
Postdoctoral Fellowship for Young Scientists.  This research was
supported by NASA through grants NAG81027 and NAG81755 to the
University of Maryland.\\

\end{acknowledgements}

\end{document}